\documentclass[ %
12pt,
a4paper,
ngerman,
english,
]{scrartcl} %

\usepackage{babel}              
\usepackage[utf8]{inputenc}   
\usepackage[T1]{fontenc}        
\usepackage[scaled]{helvet}     
\setcapindent{0em}

\usepackage{amsmath}
\usepackage{amsfonts}
\usepackage{amssymb}
\usepackage{amsthm}    
\usepackage{graphicx}   
\usepackage{caption}
\usepackage{subcaption}
\usepackage[automark]{scrpage2} 
\usepackage{geometry} 
\usepackage{color}
\usepackage{epstopdf}           
\usepackage{fancybox}           
\usepackage{tabu}
\usepackage{hhline}
\usepackage[draft]{hyperref}
\hypersetup{final}
\usepackage{cleveref}
\usepackage{makeidx}            
\usepackage{enumerate}          
\usepackage{float}
\usepackage{doi}
\usepackage{bm}
\makeindex

\begin{document}

\title{Probabilistic One-Dimensional Inversion of Frequency-Domain Electromagnetic Data  Using a Kalman Ensemble Generator}

\author{Christin Bobe
  \thanks{ %
    Corresponding Author: christin.bobe@ugent.be, %
    Department of Environment, %
    Ghent University, %
    Gent, %
    Belgium %
  } %
\and Ellen Van De Vijver
\footnotemark[1] %
\and Johannes Keller%
  \thanks{ %
    Institute for Applied Geophysics and Geothermal Energy, %
    RWTH Aachen University, %
    Aachen, %
    Germany %
  } %
\and Daan Hanssens 
\footnotemark[1] %
\and Marc Van Meirvenne 
\footnotemark[1] %
\and Philippe De Smedt 
\footnotemark[1] %
}%

\maketitle

$\copyright$ 2019 IEEE.  Personal use of this material is permitted.  Permission from IEEE must be obtained for all other uses, in any current or future media, including reprinting/republishing this material for advertising or promotional purposes, creating new collective works, for resale or redistribution to servers or lists, or reuse of any copyrighted component of this work in other works.

\newpage

\begin{abstract}
  Frequency-domain electromagnetic (FDEM) data of the subsurface are
  determined by electrical conductivity and magnetic susceptibility. %
  We apply a Kalman Ensemble generator (KEG) to one-dimensional
  probabilistic multi-layer inversion of FDEM data to derive
  conductivity and susceptibility simultaneously. %
  The KEG provides an efficient alternative to an exhaustive Bayesian
  framework for FDEM inversion, including a measure for the
  uncertainty of the inversion result. %
  Additionally, the method provides a measure for the depth below
  which the measurement is insensitive to the parameters of the
  subsurface. %
  This so-called depth of investigation is derived from ensemble
  covariances. %
  A synthetic and a field data example reveal how the KEG approach can
  be applied to FDEM data and how FDEM calibration data and prior
  beliefs can be combined in the inversion procedure. %
  For the field data set, many inversions for one-dimensional
  subsurface models are performed at neighbouring measurement
  locations.  %
  Assuming identical prior models for these inversions, we save
  computational time by re-using the initial KEG ensemble across all
  measurement locations. %
\end{abstract}

\section{Introduction}
Exploring the subsurface electrical conductivity (EC) and magnetic
susceptibility (MS) is interesting to geophysicists as anomalies in
both quantities can often be associated with resources, geological
structures, contamination or human activity. %
Frequency-domain electromagnetic (FDEM) measurements are determined by
the EC and MS of the subsurface. %
In recent years, new applications of FDEM measurements have been
explored and the design of FDEM has been tailored to facilitate survey
practice. %
These developments have enabled surveying large areas efficiently
(e.g., \cite{von2014three} and \cite{de2013exploring}), and surveying
in highly conductive and/or magnetic environments (e.g.,
\cite{delefortrie2014frequency} and \cite{simon2015mapping}). %
Inversions of such data sets may require large computational effort
for two reasons: the size of the data set itself, and the
non-linearity of the forward model exceeding the conditions of the
low-induction number approximation \cite{mcneill1980electromagnetic}
requiring non-linear inversion methods. %
To tackle these issues, we present an efficient probabilistic
inversion method for FDEM data based on the Kalman ensemble generator
(KEG, \cite{nowak2009best}). We use a non-linear forward model that
applies a full solution of Maxwell's equations. %

The KEG presented here uses both the in-phase (IP) and out-of-phase
(OP) component of the FDEM response to invert for subsurface EC and MS
simultaneously. %
Earlier work on FDEM inversion, for example by
\cite{beard1998simultaneous}, \cite{huang2000airborne},
\cite{sasaki2010multidimensional}, and
\cite{farquharson2003simultaneous}, emphasizes the importance of using
both components of the response to receive reliable inversion results
in magnetic environments. In particular, Beard and Nyquist
\cite{beard1998simultaneous} describe how including the IP component
can avoid systematic underestimation of EC when a significant IP shift
is recorded. %

The aforementioned publications apply classical, deterministic
inversion approaches yielding a single model parameter realisation
\cite{aster2005parameter}. %
We propose a pragmatic approach to probabilistic FDEM inversion by
applying the ensemble-based KEG. %
The KEG method can be seen as a trade-off between two extreme
approaches to inverse problems: (1) using a deterministic inversion
that quickly finds a single model satisfying the data, and (2)
computing a large number of possible parameter models in search
techniques aiming to be exhaustive (e.g., Markov-chain Monte-Carlo
methods \cite{minsley2011trans}). %

Ensemble-based inversions have been proven to be efficient and robust
\cite{nowak2009best}, but at the same time comprehensive enough to
characterize the uncertainty of the result. %
In the context of Kalman methods, uncertainty is characterized by
standard deviations (STD) of model parameters, which are a measure for
the parameter spread of the models that match the field
measurements. %
The KEG method uses an equation equal to the update step of the
Ensemble Kalman Filter \cite{evensen2003ensemble}, but it performs
updates exclusively in the model parameters \cite{nowak2009best}. %


The novelty of our work lies in the application of the KEG to the
inversion of FDEM data. We update prior EC and MS ensembles
simultaneously, based on the measurements of the IP and OP component
of the secondary electromagnetic field. %
Additionally, it is shown how correlations computed in the KEG can be
used as a proxy for the measurement sensitivity. %
Using this sensitivity proxy, we determine a depth of negligible
sensitivity, the so-called depth of investigation of a particular
measurement setup. %
The KEG provides an interesting inversion method for geophysical data,
especially when moving towards large and multi-dimensional forward
models. %

The application of the KEG is demonstrated for multi-configuration
FDEM data: first, on synthetic data including vertical variation in
both EC and MS, and second, on a field data set from an archaeological
site in Dorset, United Kingdom. %
The field data was collected with a small-loop FDEM device consisting
of one transmitter and several fixed receiver coils rigidly installed
on a mobile sled system. %
This way, all data points in the field data set were acquired using an
identical measurement setup. %
Our results show that the KEG can be used for simultaneous inversion
of EC and MS. %
We show how the method delivers model uncertainties. %
Additionally, it is demonstrated that, if the same prior model is
assumed for multiple data points, a single prior ensemble can be
re-used in the inversions at these multiple data points. %

\section{Methodology} \label{sec:theory}

\subsection{Forward model}
\label{sec:fwd}

A popular measurement setup for FDEM data are so-called loop-loop
systems, which are characterized by the usage of one transmitter coil
and one or multiple distinct receiver coils \cite{everett2013near}. %
The transmitter coil is excited by an alternating current. %
It generates the primary electromagnetic field which propagates into
the subsurface and induces alternating eddy currents in conducting
material. %
These eddy currents generate a secondary magnetic field which is
detected by one or multiple receiver coils. %
FDEM data are often expressed in terms of the in-phase (IP) and 90
degrees out-of-phase (OP) components with respect to the primary
electromagnetic field. %
For low-frequency applications, applying a quasi-static approximation,
these components are mainly influenced by electrical conductivity (EC)
and magnetic susceptibility (MS), whereas dielectric permittivity is
negligible. %

We compute the forward model response according to Maxwell's equations
for a one-dimensional, horizontally layered half-space. %
This forward model accounts for vertical variation of EC and MS
(Figure \ref{fig:prior}). %
As input for the computation of the FDEM measurement response, we
choose a certain coil configuration (geometry, frequency of the
primary field and transmitter moment) and subsurface realization
(discretization and electromagnetic properties), in which the deepest
layer is assumed to extend to infinite depth
\cite{hanssens2019frequency}. %

For the horizontal co-planar (HCP) coil configurations, the magnetic
field $\mathbf{H}_{ZZ}$ at the receiver coil is expressed by the
following equation \cite{ward1988electromagnetic}: %

\begin{footnotesize}
\begin{equation}
  \mathbf{H}_{ZZ}=\frac{m}{4\pi}\int_{0}^{\infty}\left[ \textrm{e}^{u_0 (z+h^{tx})} 
    + r_{TE} \textrm{e}^{u_0(z-h^{tx})}\right] \frac{\lambda^3}{u_0} 
  \mathbf{B}_0(\lambda \mathbf{r}) \textrm{d}\lambda
\label{eq:HCP}
\end{equation}
\end{footnotesize}

where $m$ is the transmitter moment, $h^{tx}$ is the height of the
transmitter coil above ground, $\mathbf{r}$ is the separation of
transmitter and receiver coil, $\lambda=(k^{2}_x+k^{2}_y)^{1/2}$ is
the horizontal wave number; whereby $u_i=(\lambda^2-k_i^2)^{1/2}$, in
which
$k_i=(\omega^2 \mu_i \epsilon_i - j \omega \mu_i \sigma_i)^{1/2}$ is
the wave number of the ith layer, with the angular frequency $\omega$;
$\sigma_i$ the conductivity, $\mu_i$ the magnetic permeability, and
$\epsilon_i$ the dielectric permittivity of the layer $i$; $j$ is the
imaginary unit $j^2=-1$; $\mathbf{B}_0$ is the Bessel function of
zeroth order; $r_{TE}$ is the reflection coefficient calculated for a
layered medium by the recursive formula given by Ward and Hohmann
\cite{ward1988electromagnetic}. %
We perform the Hankel $\mathbf{J}_0$ and $\mathbf{J}_1$ transforms
using the digital linear filter as described by Guptasarma and Singh
\cite{guptasarma1997new}. %
Formulas for the magnetic field of other dipole orientations analogous
to equation \ref{eq:HCP} are provided by Ward and Hohmann
\cite{ward1988electromagnetic}. %

FDEM data $\mathbf{d}(\omega)$ are presented in parts-per-million
(ppm) of the primary field \cite{minsley2011trans}:

\begin{footnotesize}
\begin{equation}
\mathbf{d}(\omega)=\frac{\mathbf{H}(\omega)-\mathbf{H}^0(\omega)}{\mathbf{H}^0(\omega)}\cdot 10^6
\end{equation}
\end{footnotesize}

where $\mathbf{H}$ is the total magnetic field (e.g., equation
\ref{eq:HCP}) and $\mathbf{H}^0$ is the magnetic field of free
space. %

\subsection{Bayesian parameter estimation in inverse problems}

In the next sections we will motivate our usage of the Kalman ensemble
generator (KEG) as an approximate solution of the Bayesian parameter
estimation problem. %
The Bayesian parameter estimation problem
\cite{allmaras2013estimating} consists of finding the so-called
posterior probability distribution $\rho(\mathbf{m} \mid
\mathbf{d})$. %
Here, $\mathbf{m} \in \mathbb{R}^{n_{m}}$ is a vector containing
random variables for each of the $n_{m}$ parameters to be estimated. %
In this study, $\mathbf{m}$ includes subsurface EC and MS for all
discretized subsurface layers. %
The vector $\mathbf{d} \in \mathbb{R}^{n_{d}}$ contains the $n_{d}$
observed random variables, on which the estimation is conditioned. %
The forward model from the previous section will be called $g$ 
from now on. %
The forward model $g$ allows to calculate a set of simulated
measurement data $\mathbf{d}_{sim}=g(\mathbf{m})$. %
Exact knowledge of the true physical parameters and a perfect forward
model would yield $\mathbf{d}_{obs}=\mathbf{d}_{sim}$, where
$\mathbf{d}_{obs}$ are the means of unbiased measurements. %
The posterior distribution $\rho(\mathbf{m} \mid \mathbf{d})$ is 
given by Bayes' theorem

\begin{footnotesize}
\begin{equation}
\rho(\mathbf{m} \mid \mathbf{d})= \kappa \cdot \rho_M(\mathbf{m}) \rho_D(\mathbf{d} \mid \mathbf{m}),
\label{eq:PosteriorPDF}
\end{equation}
\end{footnotesize}

with a normalization constant $\kappa$
\cite{allmaras2013estimating}. %
In the inversion, we use $\rho_D(\mathbf{d} \mid \mathbf{m})$, the
likelihood of a certain set of measurements $\mathbf{d}$ given a set
of parameters
$\mathbf{m}$. 
Computing this likelihood, the simulated measurements
$\mathbf{d}_{sim}$ are compared to the actual measurements
$\mathbf{d}_{obs}+\epsilon$, which are assumed to be unbiased with
random measurement error $\epsilon$. %
The prior information on the parameters is collected in the
probability density function (PDF) $\rho_M(\mathbf{m})$. %
Below, incorporation of prior information will be further discussed. %

\subsection{Least-squares}

Inverse problems are often solved applying variations of the
least-squares approach. %
Likewise, the KEG can be motivated starting from a probabilistic
least-squares approach. %
Readers familiar with the latter might want to skip the following
section.

A derivation of least-squares as the solution of the Bayesian
parameter estimation problem for Gaussian probability distributions
and independent measurements is given by Tarantola
\cite{tarantola2005inverse}, Chapter 3. %
Gaussian prior information can be given as

\begin{footnotesize}
\begin{equation}
\rho_M(\mathbf{m})=\textrm{const} \cdot \textrm{exp}(-\frac{1}{2}(\mathbf{m}-\mathbf{m}_{prior})^T \mathbf{C}^{-1}_M (\mathbf{m}-\mathbf{m}_{prior})),
\label{eq:rhoM}
\end{equation}     
\end{footnotesize}             

stating that the model $\mathbf{m}$ is a sample of the Gaussian
prior with mean $\mathbf{m}_{prior}$ and covariance matrix
$\mathbf{C}_M$. %
The Gaussian PDF for the measurement variables $\mathbf{d}$ is of the
same form as equation \ref{eq:rhoM} with mean $\mathbf{d}_{obs}$ and the 
corresponding covariance matrix of random observation errors $\mathbf{C}_D$. %
For measurements with Gaussian noise and parameters $\mathbf{m}$ with
Gaussian PDFs, a linear forward model $g$ will lead to a Gaussian
posterior PDF. %
The further the relation $\mathbf{d}=g(\mathbf{m})$ deviates from being
linear, the further the posterior PDF deviates from being
Gaussian. %

For nonlinear problems, Tarantola \cite{tarantola2005inverse} linearizes the forward 
function around $\mathbf{m}_{prior}$ and approximates:

\begin{footnotesize}
\begin{equation}
g(\mathbf{m}) \simeq g(\mathbf{m}_{prior})+D g(\mathbf{m}-\mathbf{m}_{prior}),
\end{equation}
\end{footnotesize}

where $D g$ is the Jacobian matrix of $g$ at $\mathbf{m}_{prior}$. %
The Gaussian posterior is defined by its posterior mean 
$\mathbf{\widetilde{m}}$ and covariance $\mathbf{\widetilde{C}}_M$:

\begin{footnotesize}
\begin{equation}
\mathbf{\widetilde{m}} \simeq \mathbf{m}_{prior}+\mathbf{C}_M D g^T (D g \mathbf{C}_M D g^T +\mathbf{C}_D)^{-1} 
\cdot(\mathbf{d}_{obs}-g(\mathbf{m}_{prior}))
\label{LSmean}
\end{equation}
\end{footnotesize}

and

\begin{footnotesize}
\begin{equation}
\mathbf{\widetilde{C}}_M \simeq (D g^T \mathbf{C}_D^{-1} D g + \mathbf{C}_M^{-1})^{-1}.
\label{LSCov}
\end{equation}
\end{footnotesize}

As stated by Tarantola \cite{tarantola2005inverse}, equation 
\ref{LSmean} and \ref{LSCov} are equivalent to the sequentially applied 
update equations of the Kalman Filter \cite{kalman1960new} as derived 
for linear stochastic systems. %

\subsection{Kalman ensemble generator}

The Kalman ensemble generator is a parameter estimation algorithm 
based on the update equation of the Ensemble Kalman Filter (EnKF, 
\cite{evensen2003ensemble}). %
The EnKF is a widely used data assimilation method. %
It was introduced as a computationally more efficient alternative to
the classical Kalman Filter \cite{kalman1960new}, which is a
sequential application of least squares. %
The EnKF uses an ensemble of states and parameters which is updated
using the so-called Kalman formula (equation \ref{eq:Kalman} as
implemented). %

The efficiency of the EnKF, and thus KEG, comes from the approximation 
of all Gaussian PDFs by an ensemble. %
An ensemble consists of a number of random samples drawn from a PDF,
the so-called ensemble members. %
In this sense, the EnKF is a Monte Carlo implementation of the Bayesian 
update problem \cite{evensen1994sequential}. %
Covariance matrices are replaced by 
sample covariances and the approach can be understood as an 
ensemble-based approach to Kalman filtering. %

The EnKF has been used in state estimation (e.g., \cite{evensen1994sequential}), parameter estimation, and 
mixtures of both (e.g., \cite{hendricks2008real}). %
If the Gaussian assumption for all involved PDFs is 
applicable, the EnKF can be used for parameter and state
estimations in nonlinear systems \cite{zhou2011approach}. %
The stationary parameter estimation approach has been described by
Nowak \cite{nowak2009best} and called the Kalman ensemble generator
(KEG) to differentiate it from the classical Ensemble Kalman
approaches. %

The prior ensemble of the KEG is a collection of randomly
drawn samples from the prior PDF. %
If the prior PDF is Gaussian as in equation \ref{eq:rhoM}, 
Gaussian random samples can be determined by a prior mean 
and the covariance matrix (see next section and Figure 
\ref{fig:prior}). %
More general prior PDFs are possible, but it has to be kept in mind 
that the derivation of the equations for least-squares  
parameter estimation makes use of the Gaussianity of the PDFs. %
The total number of samples is called $n_{ens}$. %
All samples are collected in the ensemble matrix
$\mathbf{A} \in \mathbb{R}^{n_{m}\times n_{ens}}$. %

Given $m$ discrete subsurface layers, the number of estimated
parameters for EC and MS is $n_{m} = 2m$. %
These subsurface parameters are treated as random variables. %
The prior ensembles are updated using the equation 
\cite{evensen2003ensemble}:

\begin{footnotesize}
\begin{equation}
\mathbf{A}^{Update}=\mathbf{A}+\mathbf{A'G'}^{T}(\mathbf{G'G'}^{T}+\mathbf{EE}^{T})^{-1}(\mathbf{D-G}),
\label{eq:Kalman}
\end{equation}
\end{footnotesize}

where $\mathbf{A}^{Update} \in \mathbb{R}^{2m \times n_{ens}}$ is the
posterior ensemble matrix, the update of the prior ensemble matrix
$\mathbf{A} \in \mathbb{R}^{2m \times n_{ens}}$. %
$\mathbf{A}' \in \mathbb{R}^{2m \times n_{ens}}$ is the ensemble
matrix $\mathbf{A}$ after substracting the mean value of each column
from this column. %
$\mathbf{D} \in \mathbb{R}^{n_{coils} \times n_{ens}}$ is the data
matrix containing an ensemble of FDEM measurements. %
The ensembles in $\mathbf{D}$ are generated from Gaussian PDFs with
the IP and OP measurement as mean and measurement noise as standard
deviation. %
$\mathbf{E} \in \mathbb{R}^{n_{coils} \times n_{ens}}$ is the data
matrix $\mathbf{D}$ after substracting the mean value of each column
from this column. %
$\mathbf{E}$ can be derived from the covariance matrix of random 
observation errors by Cholesky decomposition 
$\mathbf{C}_D=\mathbf{E} \cdot \mathbf{E}^T$. %
In the response matrix
$\mathbf{G} = g(\mathbf{A}) \in \mathbb{R}^{n_{coils} \times
  n_{ens}}$, we collect the responses of the samples in
$\mathbf{A}$. %
In contrast to $D g$ from least-squares, no
linearization around $\mathbf{m}_{prior}$ is used in the computation
of $\mathbf{G}$. %
$\mathbf{G'}$ is $\mathbf{G}$ after substracting the mean value of
each column from this column. %

The mean values of the samples in $\mathbf{A}^{Update}$ are considered
as the best-fit-solution to the measurement data, with their standard
deviation representing the uncertainty of this fit. %
For Gaussian PDFs and a linear forward model, the best fit solution is
equivalent to the Maximum A Posteriori estimate (MAP,
\cite{aster2005parameter}). %
In contrast to classical inversion and similar to other Bayesian
methods, the KEG allows to derive an inversion result without
introducing further regularization to the processing. %
This is possible because a regularization is implicitly introduced by
the determination of parameters through their PDFs. %
For Gaussian PDFs, this implicit regularization can be shown to
correspond to least-squares with Tikhonov regularization
\cite{aster2005parameter}. %
An overview of the inversion approach is shown in Figure
\ref{fig:flowchart}. %

In contrast to the least-squares approach from the previous section,
where mean equation \ref{LSmean} and covariance equation \ref{LSCov}
solve the inversion, for the KEG, one matrix update equation is used
to derive posterior mean and covariance (equation \ref{eq:Kalman}). %
Equation \ref{eq:Kalman} can be understood as $n_{ens}$ least-squares
computations. %
For Gaussian PDFs, a linear forward model, and infinitely large
ensembles, the KEG is equivalent to the least-squares approach. %
The KEG has two advantages over equation \ref{LSmean} and
\ref{LSCov}. %
First, all covariances are computed from the ensembles, which is
fast. %
Second, the KEG does not linearize $g$ around $\mathbf{m}_{prior}$. %
Even though a linearization is still implicit in the
least-squares-like update step, calculating the full forward model
reduces the error caused by non-linearities in the misfit
$(\mathbf{D}-\mathbf{G})$ and the calculation of covariances using
$\mathbf{G}'$. %
Assuming not too strong non-linearities around the chosen
$\mathbf{m}_{prior}$, we derive the uncertainty of the result at much
reduced cost as compared to the commonly applied MCMC methods (see
following paragraph). %

\paragraph{Comparison to MCMC}
We compare the KEG approach to well-established Markov chain-Monte
Carlo (MCMC) methods (e.g., \cite{tarantola2005inverse} and
\cite{aster2005parameter}). %
An advantage of MCMC methods is that they can handle general
non-linear inverse problems and converge to the exact solution of the
parameter estimation problem in the limit of an infinite number of
samples. %
Additionally, MCMC methods can be used for non-Gaussian PDFs. %
A drawback of MCMC methods is their large computational expense. %
They require an exhaustive search of the model space typically
applying a Metropolis- or Gibbs-sampler. %
The results of the sampler form a Markov chain in which consecutive
samples are correlated. %
This process has a lower efficiency than the KEG for three reasons:
(1) the MCMC forward model runs cannot be parallelized, (2) the
acceptance rate (for the Metropolis-sampler) is
ideally between 30 and 50 $\%$ \cite{tarantola2005inverse}, and (3) to
obtain uncorrelated samples only every $n$-th accepted sample can be
used for the posterior distribution \cite{gelman2013bayesian}. %
To highlight the importance of reducing the number of forward model
runs, we take a look at the application of the KEG shown in section
"Synthetic data set". %
There, the computation of the forward models requires approximately 97
$\%$ of the overall computing time. %
For larger forward models, minimizing the number of forward model runs
could become crucial for the feasibility of probabilistic
inversions. %
The KEG provides a computationally feasible alternative. %
As a trade-off, it is only a first-order approximation to the
parameter estimation problem, meaning that only Gaussian probabilities
can be modeled. %
However, the convergence to this first-order solution is so fast that
the KEG might deliver smaller overall errors than comparable exact
methods like MCMC approaches, especially for small CPU budgets. %

\subsection{Prior model}
\label{sec:prior}

The selection of a prior model incorporates prior knowledge into the
parameter estimation. %
This selection is important since the prior restricts the search space
of parameter models. %
This is especially true for the KEG since the KEG is restricted to
Gaussian parameter models. %
An underestimation of prior uncertainty could lead to best fits that
lie far from the true parameter values, thus rendering the standard
deviation (STD) meaningless. %
As for general probabilistic inversions, the results of the KEG have
to be analyzed with these subtleties in mind. %

\begin{figure}
\centering
\includegraphics[scale=0.87]{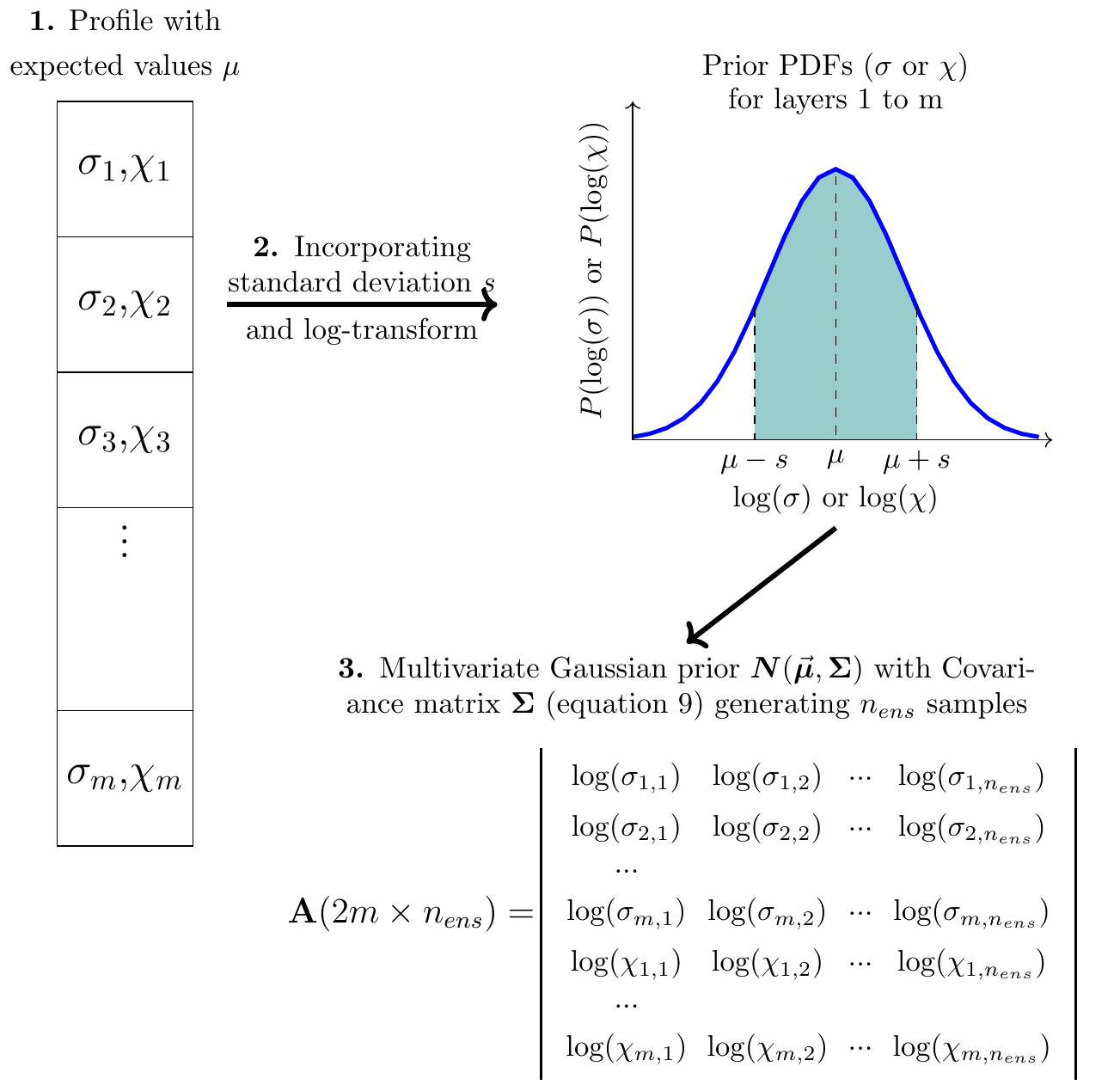}
\caption{Creating the prior ensembles for ensemble matrix $\mathbf{A}$: 
Expected EC $\sigma$ and MS $\chi$ values are initialized for a profile 
with $m$ layers. Together with the standard deviation $s$ of fluctuations 
around these expectations, a multivariate Gaussian distribution is defined. 
Input ensembles are created through selecting $n_{ens}$ samples from this 
distribution.}
\label{fig:prior}
\end{figure}

\begin{figure}
\centering
\includegraphics[scale=0.85]{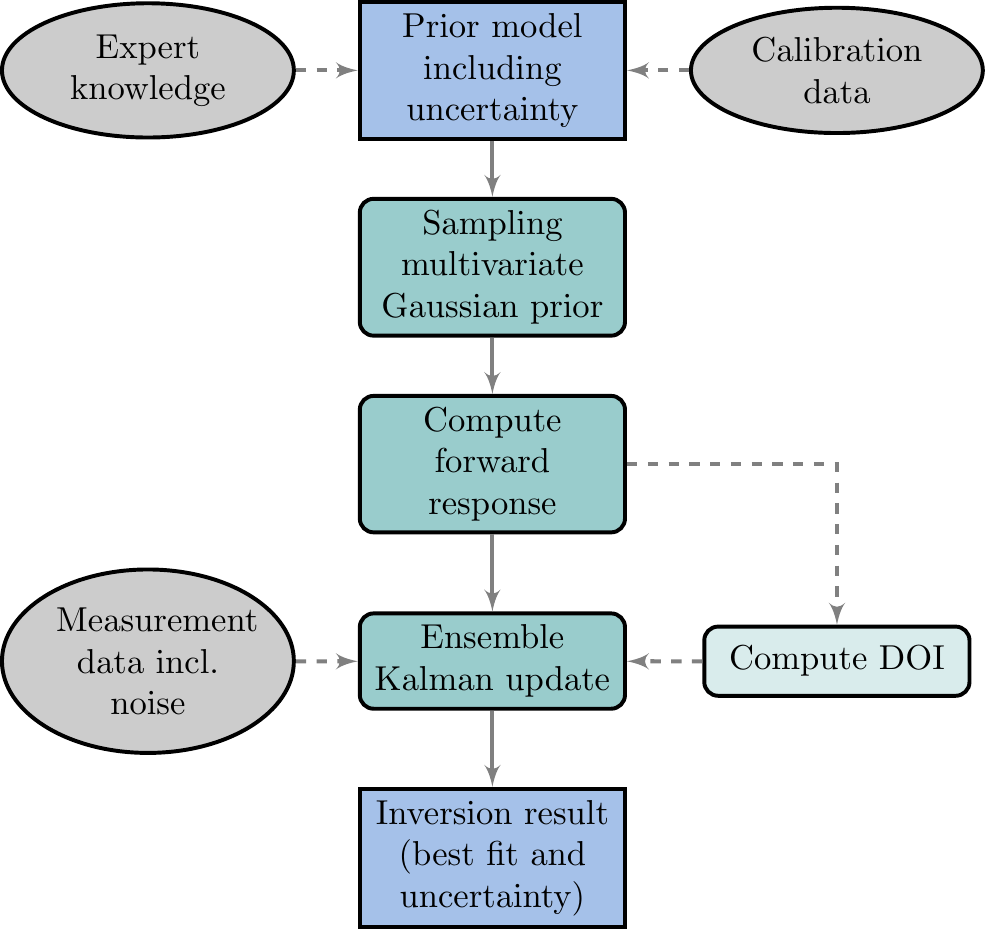}
\caption{Flow chart of the KEG inversion: Starting from the 
prior estimation of the 
model parameters, a prior ensemble (Fig. \ref{fig:prior}) is generated 
by sampling the EC and MS multivariate Gaussian prior. Correlation of 
prior and corresponding forward response ensemble can be used to compute 
the depth of investigation (DOI). The forward responses to the prior 
ensemble and the measurement ensemble are used to create an Ensemble 
Kalman update of the prior model. This update gives the posterior 
model: the inversion result.} 
\label{fig:flowchart}
\end{figure}

In later chapters, we use vertically layered, one-dimensional model
discretizations for the computations of the FDEM model. %
In this case, prior models for EC and MS are defined for each
discretization layer. %
The prior should in principle be independent of the measured FDEM
data. %
For each discretization layer, we use Gaussian prior guesses for EC
and MS consisting of a mean and a STD for log-transformed EC or MS
(Figure \ref{fig:prior}). %
The logarithm enforces positive values for EC and MS estimations,
diamagnetic effects are neglected. %
The mean of the prior ensemble serves as expected value, the STD can
be normalized by the mean to yield the coefficient of variation of
log-transformed EC or MS. %
If model parameters are expected to be correlated, a multi-Gaussian
prior can be defined through a mean vector and a covariance matrix. %

The number of discretization layers is fixed throughout the inversion
procedure and needs to be determined based on a trade-off between 
the computational cost, bias and the vertical model variability. %
A coarse discretization reduces the computational expense needed. %
In contrast, choosing a relatively finer discretization has two main 
advantages: (1) when the number of discrete layers is much larger than 
the number of expected subsurface layers, the influence of the discrete 
layer boundaries on the inversion result is reduced, and (2) a large 
number of inversion parameters entails only weak regularization and 
therefore reduces bias on the inversion result. %

\boldmath

The prior is defined as a multivariate Gaussian distribution
$N(\vec{\mu},\Sigma)$ \cite{bibby1979multivariate}: collecting the \unboldmath
expected values for the random parameters EC $\sigma$ and MS $\chi$ in the 
mean vector $\vec{\mu}$, and the corresponding variances and covariances in \boldmath
the covariance matrix $\Sigma$ \cite{hansen2006linear}. %

Prior model parameter correlations are represented by off-diagonal
covariances $\Sigma$ \cite{tarantola1982generalized}: \unboldmath %

\begin{footnotesize}
\begin{equation*}
\bm{\Sigma}(2m \times 2m)=
\end{equation*}

\begin{equation}
\begin{pmatrix}
\mathrm{Var}(\sigma_1) & \mathrm{Cov}(\sigma_1, \sigma_2) & \dotsb & \mathrm{Cov}(\sigma_1, \chi_{2m})\\
\mathrm{Cov}(\sigma_2, \sigma_1) & \mathrm{Var}(\sigma_2) & \mathrm{Cov}(\sigma_2, \sigma_3) & \\
\vdots & \ddots & \ddots &  \vdots\\
\mathrm{Cov}(\chi_{2m},\sigma_{1}) & \dotsb & \mathrm{Cov}(\chi_{2m},\chi_{2m-1}) & \mathrm{Var}(\chi_{2m})\\
\end{pmatrix}.
\label{eq:covarMatrix}
\end{equation}
\end{footnotesize}

In general, correlation is part of the prior model and should be
chosen in accordance with the available a priori knowledge. %
For the examples in this manuscript, we express correlation in terms
of a correlation function introduced by Gaspari and Cohn
(section 4.3 in \cite{Gaspari1999}) that approximates a Gaussian-shaped decrease of
correlation in the covariance matrix. %
This way, we compute our prior covariances using the a priori defined
STDs and a correlation length for the model parameters. %
In general, much more complicated correlation functions could be
introduced, but for the few information that are usually available
about the subsurface, STDs and a correlation length are a sufficiently
complex representation. %

For the KEG, an ensemble matrix $\mathbf{A}$ is created by collecting 
random samples from the multivariate Gaussian distribution
(Fig. \ref{fig:prior}). %
This ensemble matrix contains an ensemble of prior model realizations,
which are used as input for the forward model. %
A large ensemble size $n_{ens}$ is desirable since small ensembles can
introduce bias to the processing steps of the KEG. %

In this study, we restrict our prior models to have spatially uniform
mean and STD. %
This facilitates an evaluation of the KEG approach since the influence
of the prior is minimized. %
In general, more sophisticated prior modeling, for example using
geostatistical approaches, may improve inversion results (e.g.,
\cite{de2016structure} and \cite{hansen2006linear}).  %

\subsection{Sensitivity and depth of investigation}
\label{sec:DOI}

Using the KEG, the measurement sensitivity can be expressed in terms
of the correlation between the prior ensemble of EC or MS of a certain
layer and the corresponding forward response. %
The depth of investigation (DOI) is usually defined as the depth from
which surface data are insensitive to the investigated physical
property of the subsurface \cite{oldenburg1999estimating}. %
Thus, if variation in prior realizations below a certain depth has no
influence on the forward response the correlation should be zero. %
This is never exactly the case due to sampling uncertainty. %
But once the correlations are smaller than a certain threshold, it can be
assumed from the correlation data that the depth of investigation is
reached. %
The choice of the absolute threshold value is as arbitrary as for 
other DOI estimation methods \cite{vest2012global}. %
In any case, the threshold correlation should be chosen larger than the 
present spurious correlation resulting from the undersampling bias. %
The covariance of the prior ensembles $\mathbf{A}$ and the forward
response ensembles $\mathbf{G}$ are expressed by:

\begin{footnotesize}
\begin{equation}
\textrm{\textbf{Cov}}(\mathbf{A,G}^{T})=\mathbf{A'G'}^{T} * \frac{1}{n_{ens}-1} 
\label{eq:cov}
\end{equation}
\end{footnotesize}

with the number of ensemble members $n_{ens}$. %
Normalizing equation \ref{eq:cov} to the STDs of the
respective parameters gives the correlation of these parameters for
each layer assumed in the model. %
Since the matrix multiplication $\mathbf{A'G'}^{T}$ is part of the
KEG (equation \ref{eq:Kalman}), it is available from the general algorithm 
without further computation. %

The usage of the correlation as a sensitivity measure builds on the
computation of different prior model realizations. %
In classical sensitivity estimation approaches, sensitivities are 
derived from finite-difference approximations \cite{mcgillivray1990methods}. %
This corresponds to a local slope analysis. %
For the KEG, not a local slope, but a global variance is analyzed and is 
here interpreted as a measurement sensitivity proxy. %

\section{Synthetic data set}\label{sec:synex}

\begin{figure} 
\centering
\includegraphics[scale=0.38]{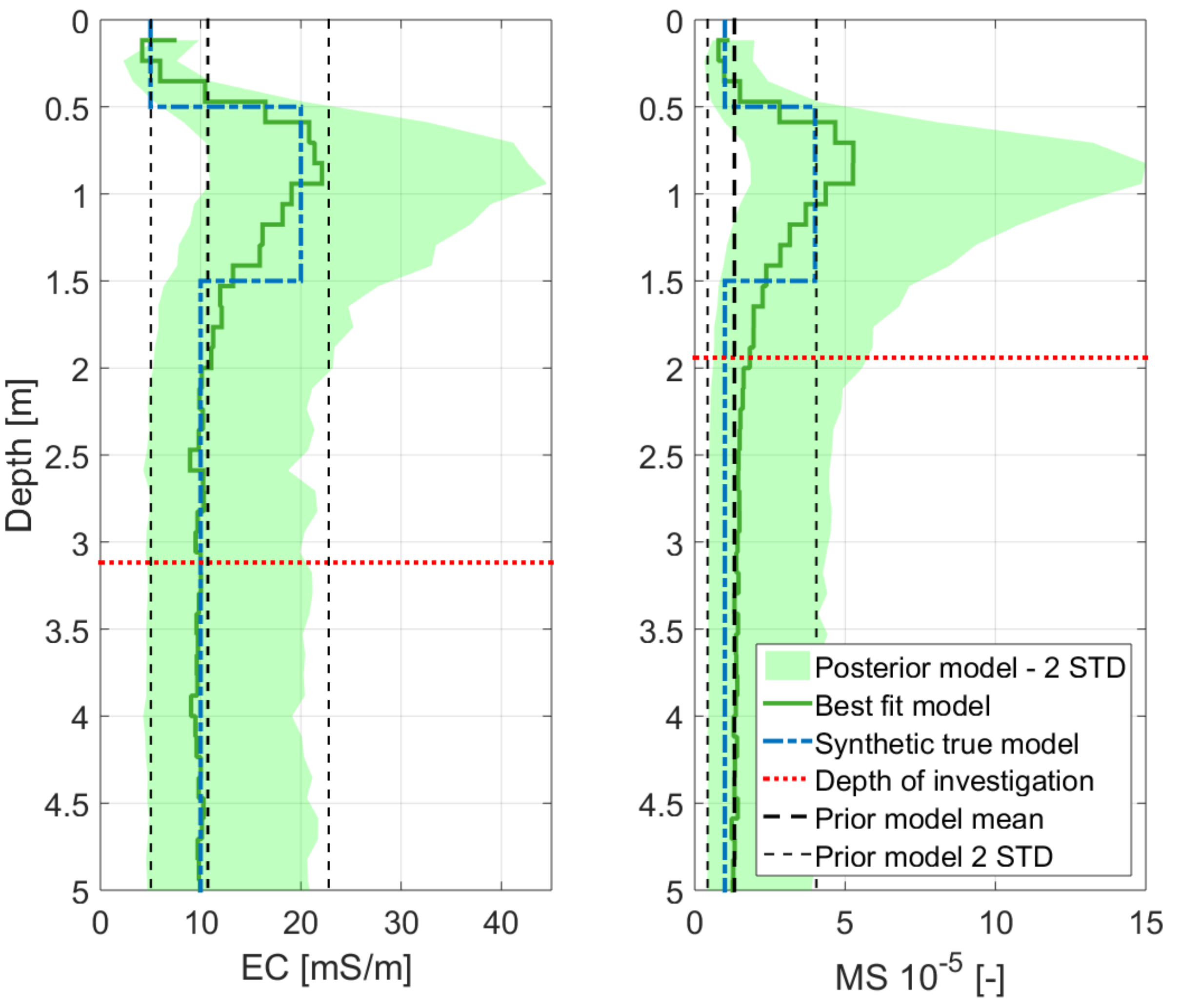}
\caption{Statistics of the posterior PDFs for the KEG inversion of
  synthetic one-dimensional model: best fit (mean) and two standard
  deviations (STD) for EC (left) and MS (right), compared to the
  synthetic true. Additionally, the mean and two STDs of the prior
  PDFs and the depth of investigation are shown.}
\label{fig:synmodel_inv}
\end{figure}

We demonstrate the KEG inversion procedure on a one-dimensional
synthetic subsurface model. %
The model includes a depth-dependency of both EC and MS realized in
three layers (Fig. \ref{fig:synmodel_inv}). %
The top layer is 50 cm thick, the intermediate layer is 1 m thick, the
bottom layer extends to infinite depth. %
EC is set to 5 mS/m for the top layer, 20 mS/m for the intermediate
layer, and 10 mS/m for the bottom layer. %
MS is set to 1$\cdot10^{-5}$ for the top layer, 4$\cdot10^{-5}$ for
the intermediate layer and 1$\cdot10^{-5}$ for the bottom layer. %

Forward model responses are simulated at four assumed receiver coils,
two coils in horizontal co-planar configuration (HCP) with 1 m and 2 m
distance to the receiver, respectively, and two coils in perpendicular
configuration (PRP) with a distance to transmitter of 1.1 m and 2.1 m,
respectively. %
The coil centers are assumed 16 cm above the ground surface. %
The transmitter moment \cite{ward1988electromagnetic} of the
transmitter coil is set to one. %
The operating frequency is set to 9000 Hz. %
These forward model parameters reflect the measurement set-up of the
field data case discussed further below. %

We choose a prior PDF consisting of Gaussian distributions. %
For each parameter (EC and MS), we assume spatial stationarity and
specify constant mean and STD throughout the model domain. %
First, we simulate synthetic measurement samples of the
logarithm-transformed synthetic true parameter values for intervals of
10 cm down to a depth of 5 m. %
These measurement samples are used as synthetic validation, thus
playing the role of vertical in-situ measurements in a field data
case. %
The mean of the prior PDF is chosen as the sample mean of the
synthetic samples. %
The sample mean is 10.7 mS/m for EC, and 1.32$\cdot10^{-5}$ for MS. %
For the same synthetic samples, the sample STD is calculated and used
as the STD for the prior Gaussian PDF. %
After an inverse log-transform, the prior STD intervals range from 7.9
mS/m to 16.8 mS/m for EC and from 0.75$\cdot10^{-5}$ and
2.8$\cdot10^{-5}$ for MS (Fig. \ref{fig:synmodel_inv}). %
For this example, there is no correlation introduced between the prior
model parameters. %
This corresponds to the choice of a correlation length that is
significantly smaller than the thickness of discretization layers. %
One motivation for this choice is the expectation of abrupt layer
boundaries. %
Consequently, off-diagonal elements of the covariance matrix are zero
(eq. \ref{eq:covarMatrix}). %
For the KEG, a prior ensemble of size 10,000 is created from the prior
PDF. %

\begin{figure*}[!t]
\centering
\includegraphics[scale=0.495]{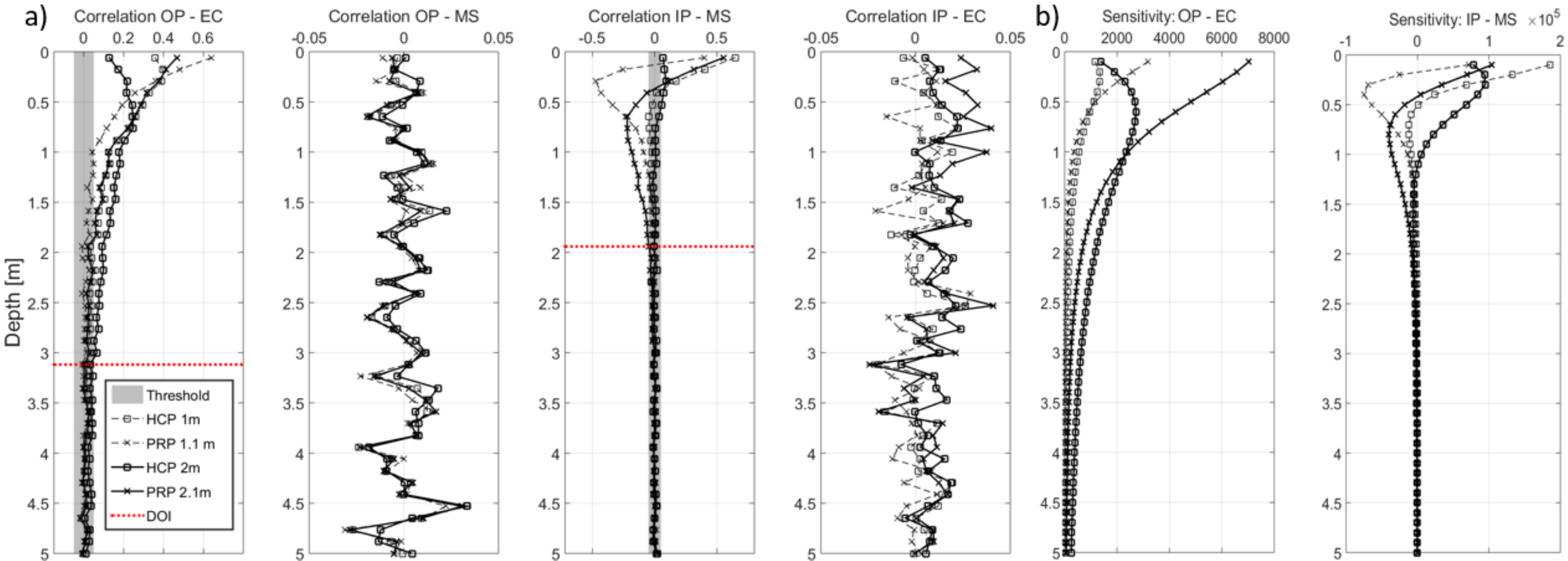}
\caption{a) Correlation between the prior ensemble of EC and MS and the four simulated 
out-of-phase (OP) and in-phase (IP) electromagnetic signals. The depth of investigation (DOI) is 
marked where all correlations fall below the threshold 0.05; b) Sensitivities of the four simulated OP signals for EC (left) and the four 
simulated IP signals for MS (right) for the receiver coils. Sensitivities are derived from perturbations of the synthetic true model.}
\label{fig:Sens}
\end{figure*}

First, we investigate the correlations between the prior parameter
ensembles (EC and MS) and the forward responses (OP and IP). %
These correlations are shown in Figure \ref{fig:Sens}a. %
The correlation between OP and MS fluctuates around zero for all four
simulated receiver coils. %
The correlation between IP and EC is similarly small, but with values
slightly larger than zero. %
Stronger correlation showing systematic variation is found for the
correlations between OP and EC, and IP and MS. %
The general shape of these two correlation functions is in agreement
with the respective differential sensitivity for the synthetic true
model (Fig. \ref{fig:Sens}b). %

As significant correlation is present only between OP and EC, and IP
and MS, we use these correlation functions for estimating the DOI of
the simulated measurement setup. %
The respective DOIs are shown in Figure \ref{fig:Sens}a. %
We set the DOI to the depth at which correlations for all coils fall
below a threshold of 0.05. %
The DOIs are at 3.12 m for EC and 1.94 m for MS. %
This is in agreement with the shallower sensitivity for IP signals
compared to OP signals as computed by perturbation of the synthetic
true model (\cite{mcgillivray1990methods}, Figure \ref{fig:Sens}b). %

For the update step, the
algorithm requires an additional ensemble of FDEM measurements. %
The STD for the synthetic measurement values was set to 0.01 ppm for both 
the OP and IP response. %

The KEG update of the prior model, is shown in Figure
\ref{fig:synmodel_inv} as best fit and corresponding posterior
uncertainty. %
For both EC and MS, the best fit detects parameter value contrasts for
the three layers present in the synthetic true model. %
Below the DOI, the best fit is a reproduction of the prior mean since
equation \ref{eq:Kalman} can be approximated by
$\mathbf{A}^{Update} \approx \mathbf{A}$ for lowly correlated
parameters. %

The uncertainty of the best fit is shown as intervals of two 
standard deviations from the mean (Fig. \ref{fig:synmodel_inv}). %
For all parameters, the synthetic true model is contained within these 
two standard deviations. %
The posterior uncertainty is smaller than the prior uncertainty in the top 
layer. %
Closer to the second layer, the best fit values as well as the uncertainty 
are larger. %
For the bottom layer, the synthetic true EC and MS are smaller 
than for the intermediate layer. %
Accordingly, the best fit renders lower values. %
The uncertainty is also decreasing from the center of the second 
layer downwards. %

To evaluate the performance of the inversion approach, we calculate 
the RMSE between the best fit of the inversion result and the 
mean of the synthetic true for the whole model down to the 
respective DOI for EC and MS. %
The RMSEs are compared to the analogous RMSEs between prior model 
and synthetic true. %
Additionally, for the best fit vector and the prior mean vector, 
the corresponding IP and OP forward responses for the four coils 
are calculated. %
For these responses, the RMSE with the synthetic true response is 
computed. %
The RMSE values for IP and OP signals are finally summed up and shown in 
Table \ref{tab:RMSE}. %

\begin{table}
\centering
\caption{Comparing the root mean square error (RMSE) of the synthetic
  ground truth with (1) mean vector of the prior ensembles and (2)
  best fit vector of the KEG inversion up to the respective DOI
  (Fig.  \ref{fig:synmodel_inv}), and the corresponding modelled
  measurement responses for in-phase (IP) and out-of-phase (OP)
  component for these respective vectors are compared to the
  synthetic-true measurement data by their respective RMSEs.}
\label{tab:RMSE}
\vspace{0.9mm}
\begin{tabular}{c|c|c|c|c|}
RMSE & EC profile & MS profile & OP-signal & IP-signal\\
\hline
Prior mean &  5.5 mS/m & 1.71$\cdot10^{-5}$ & 42.4 ppm  & 4.2 ppm \\
Best fit & 2.1 mS/m & 0.9$\cdot10^{-5}$ & 19.2 ppm & 0.7 ppm\\
\end{tabular}
\end{table}

For EC the RMSE reduces from 5.5 mS/m for the prior model to 
2.1 mS/m for the best fit. %
For MS the value reduces from 1.7$\cdot10^{-5}$ to 
0.9$\cdot10^{-5}$. %
Decreased RMSEs for the best fit can also be observed for the OP and IP 
signal. For OP the value goes from 42.4 ppm to 19.2 ppm, for 
the IP signal the RMSE is reduced from 4.2 ppm to 0.7 ppm. %
The decreased RMSEs for the best fit are in accordance with the form 
of the best fit in Figure \ref{fig:synmodel_inv}. %

The computation time of the KEG update is 35.3 seconds. %
Of this total computation time, 34.3 seconds were consumed by the
forward model runs. %
Thus, the repeated computation of forward models uses 97 $\%$ of the
overall computation time, making the forward runs the most
computationally expensive step in the inversion. %
When using a MCMC approach, we expect the number of forward model runs
to be increased at least by a factor of three compared to the KEG. %
This illustrates how the KEG benefits computational efficiency. %

\section{Field data set}\label{sec:fieldex}

An FDEM field data set was collected on a 1.3 ha area in Knowlton
(Dorset, United Kingdom). %
The local subsurface is dominated by a thin rendzina soil cover
(around 20 cm thick), developed in loessic sediments overlaying
Cretaceous chalk bedrock. %
While the topsoil is strongly magnetic (MS in the order of
1$\cdot10^{-3}$), the bedrock is diamagnetic rendering an overall
background susceptibility of zero. %
Surveyed under dry conditions, the area has a low EC, varying around 7
mS/m, while the topsoil is slightly more conductive than the bedrock
\cite{delefortrie2018low}. %
In the inversion, we aim to show the contrast in EC and MS between
those layers. %
Additionally, the area is known for several henge structures from the
Stone Age. %
These structures include man-made ditches which often cause magnetic
anomalies in the near subsurface and thus may be detected by FDEM
measurements \cite{delefortrie2018low}. %

For the FDEM measurement, a DUALEM-21S (Dualem, Canada), was used with 
simultaneous geo-referencing of the data points using a GPS device. %
The FDEM instrument configurations are identical to the one 
described in the section on the synthetic data set, i.e. two horizontal 
co-planar (HCP) and two perpendicular (PRP) coil set-ups. 

\begin{figure*}
\centering
\includegraphics[scale=0.49]{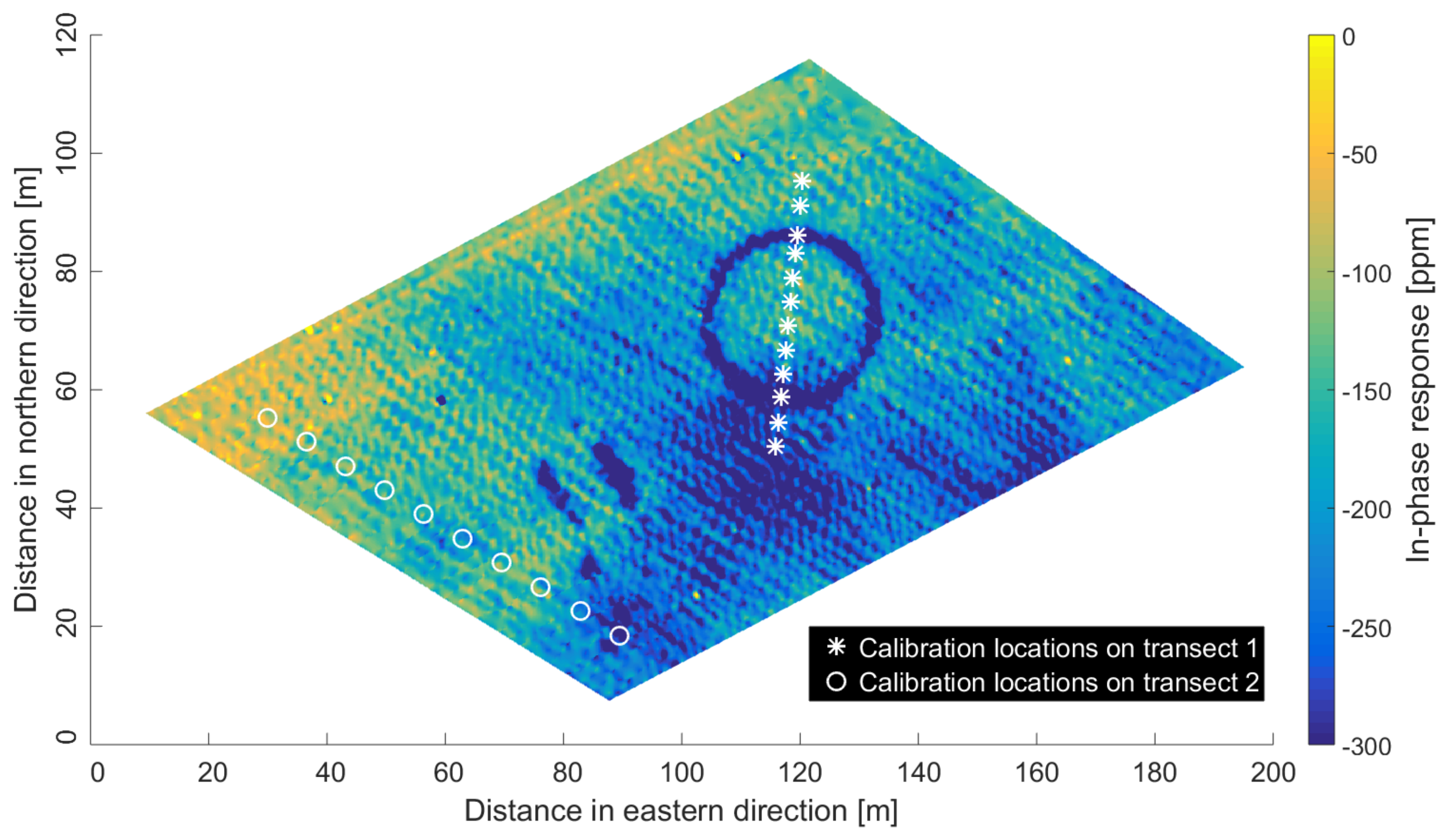}
\caption{Subset of the field measurement data: In-phase data of the PRP 1.1
  m coil, interpolated (natural-neighbour) to a 0.3$\times$0.3 m grid
  and corrected for offset error, signal drift and shift. %
  Calibration data were collected along two transects.} %
\label{fig:allIPknowlton}
\end{figure*}

\begin{figure*}
\centering
\includegraphics[scale=0.49]{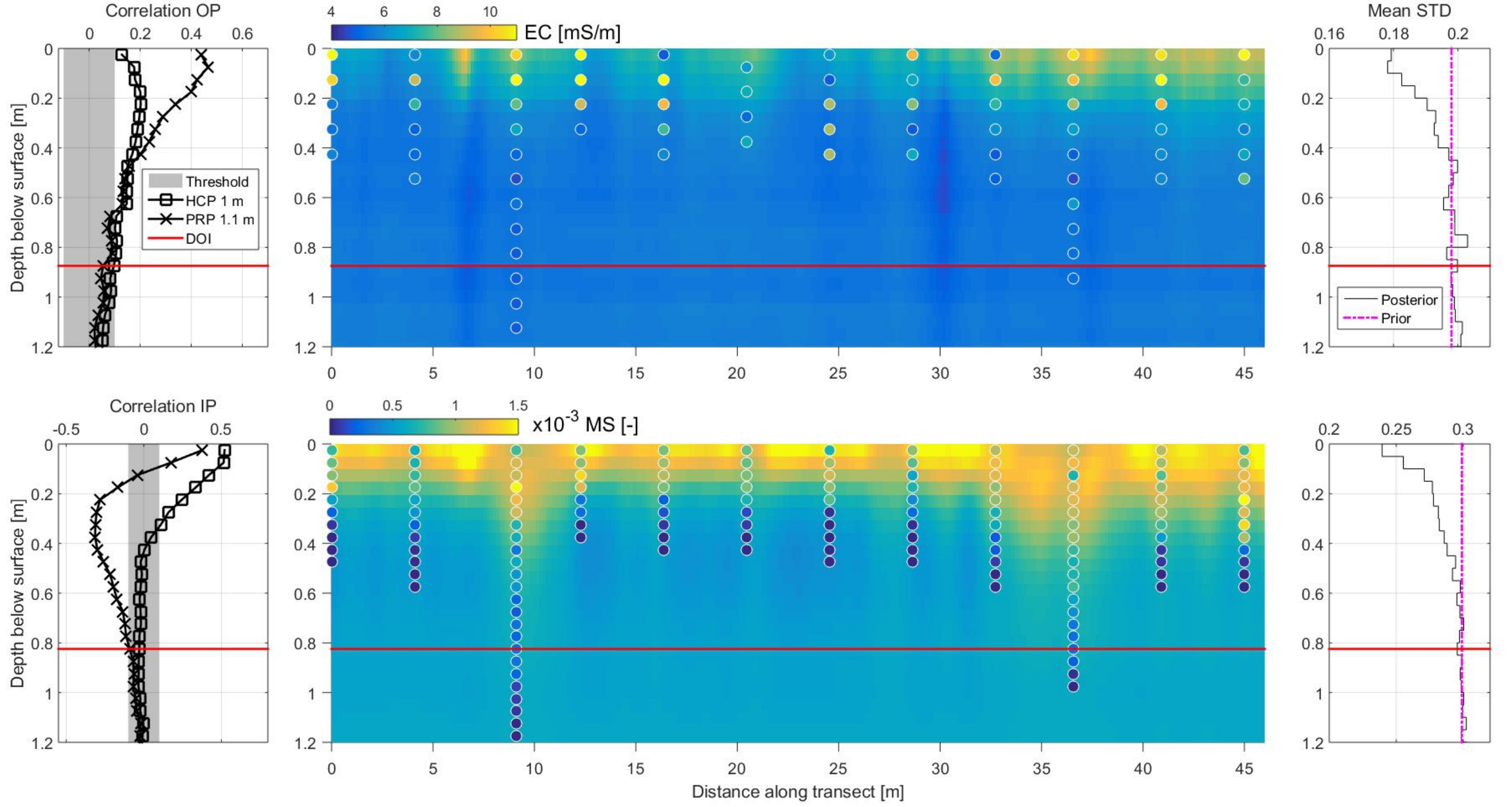}
\caption{Left: Correlation between OP electromagnetic signal and EC (top),  
correlation between IP signal and MS (bottom). The depth of investigation 
(DOI) is indicated where correlation falls below 0.1. Center: Inversion results 
(best fit) for EC (top) and MS (bottom) compared to the calibration 
data (circles with fill color corresponding to data value). Right: Standard 
deviation (STD) of the prior PDFs and mean STD of the posterior PDFs for 
logarithmic EC (top) and logarithmic MS (bottom).}
\label{fig:valid}
\end{figure*}

FDEM data were collected along parallel lines in southwest-northeast
direction, with a distance of approximately one meter, covering the
entire survey area. %
Additionally, vertical profiles of EC and MS values were collected at
22 locations in the survey area, distributed along two transects 
(Fig. \ref{fig:allIPknowlton}). %
Twelve vertical profiles are located along transect 1 crossing 
the man-made ditch structure in the center of the survey area. %
The remaining ten locations are positioned along transect 2 in the 
southwest of the survey area, where a large variation of the in-phase 
response is observed. %
All vertical profiles consist of measurements in intervals of 5 to 10
cm, some reaching depths of 1.2 m. %
MS data were collected with a Bartington MS2H downhole probe in a 2.5 cm 
diameter gouge borehole down to at least 15 cm in the chalk bedrock 
(i.e. repeated recording of diamagnetic bedrock response). %
The EC measurements were carried out with a UMP-1 BTim field probe (UGT) in 
a 5 cm diameter borehole, prepared with a riverside corer at 5 to 10 
cm increments. %
For most vertical profiles, the calibration data shows the highest 
EC (mean of 9.5 mS/m) and MS (mean of 1$\cdot10^{-3}$) in the top layer, 
and lower EC (mean of 5 mS/m) and zero MS at greater depth 
(Figure \ref{fig:valid}). %
This pattern of the calibration data is consistent with the geological 
considerations above. %
At the ditch locations, non-zero MS is present at depths down to 90 cm. %
Whereas, a background MS of zero is found at depths around 20 to 30 cm next 
to the ditch structure. %

Before inversion, FDEM data were processed by applying several
corrections. %
First, the data were corrected for the spatial offset between coil
midpoints and GPS recording position, as described by
Delefortrie et. al. \cite{delefortrie2016evaluating}. %
Then, the data were drift corrected using a tie line approach, as
described in \cite{delefortrie2014efficient} and \cite{de2016identifying}. %
Subsequently, the data were interpolated to a regular grid (cells of 0.3 m 
by 0.3 m) using a natural neighbour interpolation. %
The IP response of the FDEM data suffered from severe 
systematic errors (e.g., \cite{minsley2012calibration}). %
These errors were accounted for by comparing the field data to
a simulated one-dimensional forward response based on the calibration
profiles from both transects \cite{delefortrie2018low}. %
As a result, two coils were not considered in further processing. %
The data from the PRP 2.1 m coil exhibited no clear correlation with
the modelled IP forward response. %
Following Delefortrie et. al. \cite{delefortrie2018low}, we assume that the
data from this coil is heavily influenced by ploughing. 
Additionally, OP values from the HCP 2 m coil exceed the modelled OP 
response by up to an order of magnitude. %
Therefore, the data from both the PRP 2.1 m coil and the HCP 2 m
coil were excluded from the inversion processing. %
This leaves the responses of the two coils HCP 1m and PRP 1.1 m to be 
considered during the inversion. %

Data from transect 1 crossing the man-made ditch structure 
are inverted using both the OP and IP responses of the HCP 1 m and 
PRP 1.1 m coil. %
This transect was selected for two reasons: (1) it shows a relatively 
wide range of response values and lateral variation, and (2) the inversion 
results can be validated using the calibration data as ground truth. %

The prior model of the inversion is defined on 40 discretization
layers, each with a thickness of 5 cm. %
Uniform mean and STD were set for the prior model. %
The value of the mean and of the STD are derived from the sample mean
and sample STD of the calibration data from both transects (analogous
to the procedure described for the synthetic data set). %
Further, the calibration data suggest only gradual vertical changes of
EC and MS. %
For this reason, a correlation between prior model parameters was
introduced with a correlation length of 5cm which is in the order of
magnitude of the sampling interval of the calibration data. %
For the covariance matrix (eq. \ref{eq:covarMatrix}), this leads to a
correlation of approximately 0.2 for the first off-diagonals and zero
correlation for all subsequent off-diagonals. %
A forward run for 10,000 samples of the uniform prior model was
carried out initially. %
Subsequently, the correlation between prior and forward response was
computed and used as a measure for sensitivity following the approach
described above (Figure \ref{fig:valid}).  
After visual assessment of the correlation curves, the threshold
correlation for the DOI determination was set to 0.1. %
For EC, this yields a DOI of 0.87 m, and for MS a DOI of 0.82 m. %
This is in good agreement with the sensitivities as computed by the
forward model following a perturbation of the calibration data
\cite{mcgillivray1990methods}. %

The FDEM measurement ensembles for the two coils were created using the 
noise level of the instrument (20 ppm) as STD of the measurement PDFs 
\cite{dualemmanual}. %
Finally, EC and MS values were updated assimilating the measurements along 
transect 1. %

The posterior best fit profiles are shown in Figure
\ref{fig:valid}. %
Best fit EC values range from 4 to 11 mS/m. %
Best fit MS values are spread out over one order of magnitude 
(from 2$\cdot10^{-2}$ to 1.5$\cdot10^{-3}$). %
Alongside the best fit profiles, the mean STDs of the posterior PDFs (for logarithmic EC and MS values) are shown in Figure \ref{fig:valid}. %
The STDs indicate the uncertainty of the inversion result. %
Since the same prior ensemble was used for all vertical profiles, 
the obtained uncertainty is almost identical for each vertical profile along the transect. %
Therefore, only its mean value is shown in Figure \ref{fig:valid}. %
Possible variation in STD may occur due to the sampling 
differences between the measurement ensembles but this variation is small. %
The posterior STD is smaller than the prior STD (0.198 for EC and 0.299 for MS) 
at the top of the profile. %
With increasing profile depth the STD is rising until, near the 
DOI, the prior STD is reached. %

As validation of the inversion result, the best fit is compared to the 
calibration data (Fig. \ref{fig:valid}). %
The general variation in EC and MS is sufficiently recovered in the 
best fit. %
As expected, we find higher EC and MS for the upper tens of centimeter. %
Below, EC and MS values drop. %
At approximately 9.5 m and 37 m distance along the transect, the ditch 
structure is clearly visible as highly susceptible wedges. %
In accordance with the calibration profiles, the ditches 
are less distinct in the EC profile. %

Considering absolute values, MS values are systematically 
overestimated in the best fit. %
This might be explained by the severe offset errors in the raw IP signals 
which might not be sufficiently corrected by the comparison to the simulated 
forward responses of the calibration data. %

Below approximately 70 cm, MS values are close to the profile mean value. %
This mean value is close to the mean of the MS prior model 
(6.1$\cdot10^{-4}$). %
This is in agreement with the flattening of the IP correlation for the 
PRP 1.1 m coil which indicates a low sensitivity of the IP response to MS. %
Like the posterior mean, the posterior STD is close to the prior STD. %
As already observed for the synthetic data example, for lowly 
correlated profile sections, the inversion tends to reproduce the 
prior model. %

\section{Conclusion}

We apply the Kalman ensemble generator (KEG) in a probabilistic
inversion that allows simultaneous recovery of electrical conductivity
and magnetic susceptibility models from frequency-domain
electromagnetic data (FDEM). %
Correlations between prior electrical conductivity (EC) and magnetic
susceptibility (MS) ensembles and the corresponding forward response
ensembles are computed during the application of the KEG. %
We use these correlations to estimate sensitivities of the forward
responses to EC and MS. %
The depth of investigation is set by defining a threshold
correlation. %
To the best of our knowledge, this is the first time that the KEG has
been used for FDEM inversion. %
Discussing the KEG in a FDEM context is promising, especially when
moving towards larger and multi-dimensional forward models. %
As computationally expensive forward calculations are also an issue
for other types of geophysical data, we believe that interest in the
presented method might not be limited to the electromagnetic
community. %

The KEG inversion allows to express uncertainties of prior beliefs and
calibration data in a simplified Bayesian framework, as it is a
Monte-Carlo implementation of a Gaussian Bayesian update problem. %
The avoidance of an exhaustive search of the model space makes the
method more efficient than standard MCMC approaches. %
The error propagation in the algorithm provides a best fit to the
measurement data and a measure for the uncertainty around the best
fit. %
No additional regularization parameters are needed by the algorithm,
as a trade-off the approach is restricted to modeling approximately
Gaussian parameters. %

The KEG proves to be efficient, particularly when identical prior 
models are assumed at multiple inversion locations. %
In such cases, and when the measurement setup is constant, we can 
save computational time by re-using the initial KEG ensemble for 
all inversions with identical prior assumptions. %
For the presented field data example, we re-used one initial KEG 
ensemble for the processing of approximately 300 neighboring 
inversion locations across one transect. %

While in this work only prior models with a uniform mean vector 
and STD have been used, 
it is possible to define prior models with varying layer properties 
in order to model more heterogeneous subsurfaces. %
The behavior of non-uniform prior models and more sophisticated 
geostatistical prior correlations models can be investigated 
in the future. %

The KEG is capable of performing parameter 
estimations for a large number of model parameters. %
The approach can be extended by adding 
additional parameters to the estimation: systematic errors, 
location of the measurement device, and dielectric permittivity 
estimation for high frequency applications. %

\section*{Acknowledgments}

The authors thank Martin Green, Joshua Pollard and Mark 
Gillings for suggesting the field survey and their help during 
fieldwork. %
This project has received funding from the European Union's 
EU Framework Programme for Research and Innovation Horizon 
2020 under Grant Agreement No 721185: NEW-MINE (www.new-mine.eu). %
Philippe De Smedt is a Postdoctoral Fellow of the 
Research Foundation - Flanders (FWO), research grant: 
FWO13/PDO/046. %
The authors would like to thank the reviewers Wolfgang 
Nowak and Michael Zhdanov for their helpful and 
constructive comments. %

\end{document}